# Noise-induced acceleration of single molecule kinesin-1

Takayuki Ariga[1*], Keito Tateishi[1], Michio Tomishige[2], Daisuke Mizuno[3]
[1]*Graduate School of Medicine, Yamaguchi University, Yamaguchi, Japan.*
[2]*Department of Physical Sciences, Aoyama Gakuin University, Kanagawa, Japan.*
[3]*Department of Physics, Kyushu University, Fukuoka, Japan.*

(Dated: 13 September 2021)

The movement of single kinesin molecules was observed while applying noisy external forces that mimic intracellular active fluctuations. We found kinesin accelerates under noise, especially when a large hindering load is added. The behavior quantitatively conformed to a theoretical model that describes the kinesin movement with simple two-state reactions. The universality of the kinetic theory suggests that intracellular enzymes share a similar noise-induced acceleration mechanism, i.e. active fluctuations in cells are not just noise but are utilized to promote various physiological processes.

Fluctuations are ubiquitous and prominent in microscopic systems. The effects of the fluctuations on motions and/or chemical reactions are a long-studied field of nonequilibrium physics [1-8], and their relevance in biological systems is emerging as a hot topic [9-18]. Recently, direct observations *in vitro* have confirmed that motor proteins are violently shaken by thermal fluctuations [19]. The walking molecular motor, kinesin-1 (hereafter called kinesin), which carries vesicles on microtubules in cells [20], has long been proposed to utilize thermal fluctuations to make directed movements [3]. In addition to thermal fluctuations, living cells actively generate non-thermal fluctuations using energy derived from metabolic activities [10-12]. However, it is not clear whether and how these active fluctuations affect the function of kinesins in living cells.

In our previous study, we investigated the energetics of single-molecule kinesin *in vitro* [21], where the energy input/output balance of working kinesin was obtained using a novel nonequilibrium equation, the Harada-Sasa equality [22]. Whereas ~50% efficiency of kinesin has been reported at stall-force conditions [23], we observed only 20% power efficiency at working conditions and found that ~80% of the input free energy obtained from ATP hydrolysis ($\Delta\mu$) was not transmitted to cargo movement but dissipated via hidden paths [21], implying that kinesin has low efficiency [24]. However, it is hard to imagine that kinesin, which has been preserved after billions of years of molecular evolution, would be inefficient at cargo transport. Therefore, we hypothesized that kinesin is optimized for its actual working environment, *living cells*, but not necessarily *in vitro* [25]. The effects of these active fluctuations, which do not exist in experimental conditions *in vitro* but do occur in living cells [10-12], on the functions of individual motors are unknown.

In this study, to investigate the effects of active fluctuations on single kinesin molecules, we used an *in vitro* measurement system to apply actively fluctuating external forces (i.e. noise), artificially mimicking intracellular active fluctuations [Fig. 1(a)]. The results show that kinesin accelerates in response to the applied noise, especially under the application of a large average hindering force (load), indicating that kinesin is optimized to its actively-fluctuated working environment. Moreover, the acceleration was quantitatively explained with a mathematical model using independently determined parameters. Because of the universality of the theories behind the model, the noise-induced acceleration found in kinesin is widely applicable to other general enzymes in cells.

Active fluctuations within eukaryotic cells are mostly derived from actomyosins in cytoskeletal networks [9,11]. It has been reported that the distribution of myosin-generated fluctuations is heavily tailed in a manner similar to a Lévy stable distribution rather than a simple Gaussian distribution [13]. To reproduce the intracellular active fluctuations *in vitro*, we numerically generated Lévy-like stochastically fluctuating signals [26] and applied these signals as external forces via an optical trap to a probe bead that acted as a cargo carried by kinesin [Fig. 1(a), See Methods in [27]]. The position of the laser focus was controlled within $100 \pm 100$ nm from the bead center due to the technical limitation of the optical tweezers. The applied noise was therefore truncated by replacing values above the limit with an upper (or lower) limit, which we term "semi-truncated Lévy noise" [Fig. 2(a)]. Note that Lévy noise cannot be realized in physical systems in its exact mathematical form. Rather, semi-truncated Lévy noise more resembles the non-Gaussian fluctuations that have been found in various physical systems far from equilibrium, including active swimmer suspensions [14-16], actomyosin networks [13], and cultured cells [17].

Semi-truncated Lévy noise was applied to a single kinesin molecule in addition to a constant force (load) [Fig. 2(b)]. The average velocities and relative velocities, which are the ratio of the average velocity with and without added noise, are shown in Figs. 2(c) and 2(d). Each *marker* indicates the average load (-1~-5 pN), and the horizontal axis is normalized by the standard deviation (s.d.) of the fluctuations (noise) of the applied force. At each average load, the motor velocity gradually increased with the magnitude of the noise. The larger the load, the larger the increase in relative velocity due to the noise, but a smaller change was observed at low loads.

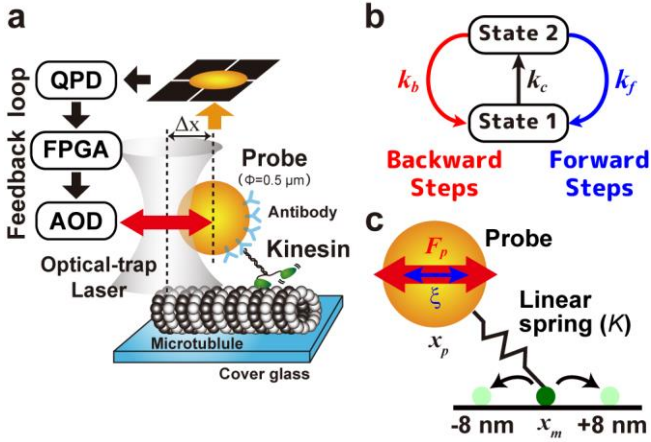

FIG. 1. (a) Schematic of the measurement system (not to scale). A single kinesin molecule is attached to an ~500 nm diameter probe particle via anti-His-tag antibody. The probe is trapped with a focused infrared laser (optical tweezers) and directed to the microtubule rail to detect kinesin movement. The probe position is obtained by projecting the image to the quadrant photodiode detector (QPD) using bright field illumination. The output voltage signal is acquired by the field programmable gate array (FPGA) board to calculate the trap position. The trap position is controlled by the output signal from the FPGA board via acousto-optic deflectors (AOD). By changing the distance between the trap and probe position ($\Delta x$: trap distance) through programming of the FPGA, an arbitrary external force, $F_p = k_{trap}\Delta x$, where $k_{trap}$ is trap stiffness, can be applied to the probe particle at an update rate of 20 kHz, which is the same as the sampling rate. Here, we applied external forces to the probe as $F_p = F_0 + F_n$, where $F_0$ is a constant force (load) and $F_n$ is a zero-mean fluctuating force (noise). (b) The two-state mathematical model for kinesin. Transitions between two internal states contain load-independent ($k_c$) and load-dependent ($k_f$ and $k_b$) transitions. $k_f$ and $k_b$ have an Arrhenius-type force dependency [Eq. (1)] that are coupled to forward and backward steps, respectively. (c) The Langevin model of the probe movement that is pulled by the kinesin molecule. The kinesin molecule is modeled as a jumping point with 8-nm back and forth steps ($x_m$), while the probe ($x_p$) is connected to the kinesin via a linear spring. The external force, $F_p$, including an average load, $F_0$, and a zero-mean fluctuating force, $F_n$, is applied to the probe, which is exposed to a thermal fluctuating force, $\xi$.

To examine the observed accelerating behavior of kinesin, we performed numerical simulations with our previously reported mathematical model [21]. The kinesin movement was modeled by the simplified kinetic model with two internal states that are connected by force-independent ($k_c$) and -dependent ($k_f$ and $k_b$) transitions [28] [Fig. 1(b)]. $k_f$ and $k_b$ have an Arrhenius-type force dependency [23]:

$$k_{\{f,b\}}(F_m) = k_{\{f,b\}}^0 \exp\left(\frac{d_{\{f,b\}}F_m}{k_B T}\right), \quad (1)$$

where $k_{\{f,b\}}^0$ is the rate constant at zero force, $d_{\{f,b\}}$ is the characteristic distance, $k_B$ is the Boltzmann constant, $T$ is the absolute temperature, and $F_m$ is the force applied to the motor. The subscripts $f$ and $b$ indicate forward and backward 8-nm steps, respectively. The simulations were conducted with Langevin dynamics of the probe, which is connected to the kinetic kinesin model via an elastic linker [Fig. 1(c)], using

$$\Gamma \frac{d}{dt}x_p = K(x_m - x_p) + F_p + \xi, \quad (2)$$

where $\Gamma$ is the viscous drag, $K$ is the spring constant of the stalk, and $x_p$ and $x_m$ are the position of the probe and motor, respectively. The external force to the probe, $F_p = F_n + F_0$, contains a constant force (load), $F_0$, and a fluctuating force (noise), $F_n$, with zero mean, and $\xi$ is white Gaussian thermal fluctuations that satisfy $\langle\xi\rangle = 0$ and $\langle\xi(t)\xi(t')\rangle = 2k_B T\Gamma\delta(t-t')$, where $\delta(t)$ is the delta function. Note that although kinesin has a nonlinear spring and we have confirmed that the spring constant depends on the external force (Fig. S2b in [27]), the numerical simulations were performed with a constant $K$ for each average load $F_0$, with $K$ measured at each constant load ($F_0$). Remarkably, despite all parameters being experimentally determined without noise (See Figs. S1,S2 and Appendix A in [27]), the simulations under semi-truncated Lévy noise [Figs. 2(e) and 2(f, markers)] show fairly similar output to the experimental conditions with noise [Fig. 2(c) and 2(d)].

Although the noise amplitudes in the above measurements are experimentally limited, simulations can be conducted at conditions exceeding the constraints. In realistic intracellular conditions, the metabolic activity of actomyosins is expected to generate much larger noise [11], where the maximum force derived from the myosin minifilament is estimated to be over 30 pN [29]. Therefore, we conducted simulations in which the noise distribution was truncated at a physiologically plausible value of 30 pN [Fig. 3(a)], indicating that the velocity at the unloaded, noiseless condition could be achieved regardless of the average load. Since the average load can be regarded as the resistance to high viscosity in cells [30], kinesin is thought to optimally utilize active fluctuations to achieve the same velocity as unloaded conditions in low viscous solutions.

The noise-induced acceleration observed here can be qualitatively explained by Jensen's inequality:

$$\langle k(F)\rangle \geq k(\langle F\rangle), \quad (3)$$

where $\langle\rangle$ indicates average, and $k$ is a convex function of $F$. Here, we regard $k$ as the kinetic rate and $F$ as the applied external force that contains both constant and fluctuating forces. As presented in Eq. (1), $k(F)$ is expressed by the Arrhenius equation, i.e. an exponential function [23]. In this case, Jensen's inequality tells us that the average of the rate constants generally increases when $F$ fluctuates. The universality of Jensen's inequality and the Arrhenius equation implies that any enzyme obeying the same Arrhenius equation can experience noise-induced acceleration.

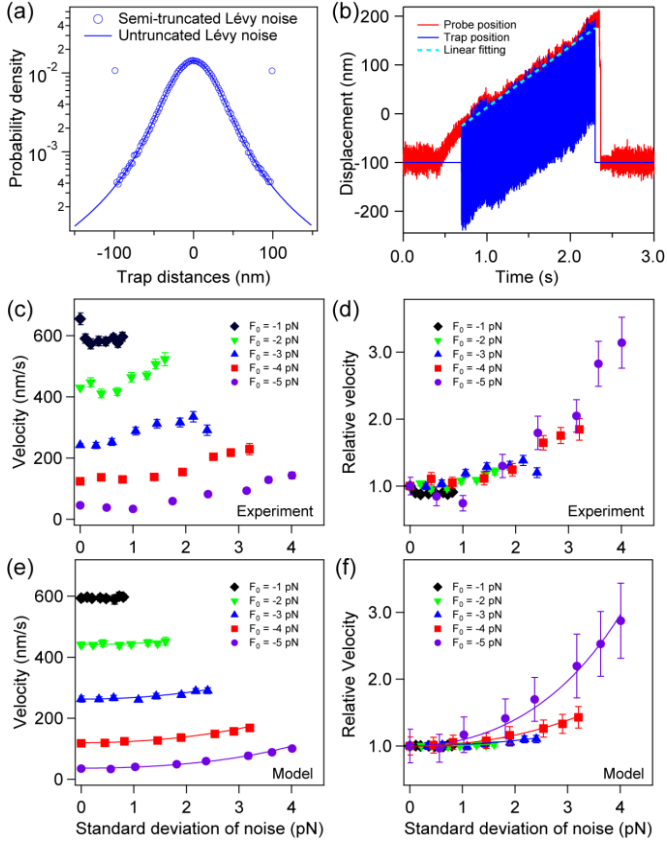

FIG. 2. (a) Distribution of semi-truncated Lévy noise. *Circles* indicate the distribution of the trap distance used in the experiments and simulations (scale parameter $\gamma = 20$ nm; see Eq. (S3) in [27]), where trap distances over ±100 nm are truncated to ±100 nm. *Line* indicates the corresponding untruncated Lévy distribution calculated by numerical integration with Eq. (S2) in [27]. (b) A typical trajectory of a probe particle pulled by a single kinesin molecule under semi-truncated Lévy noise. The external force is applied as $F_0 = -4$ pN constant force (load) by keeping the trap distance 100 nm at a trap stiffness of 0.04 pN/nm along with a mean-zero fluctuating force (noise), $F_n$, with scale parameter $\gamma = 20$ nm, which truncated the trap distance to ±100 nm. *Dotted line* indicates linear fitting of the probe trajectory to obtain the average velocity. (c) The velocity of the probe particles under semi-truncated Lévy noise of various magnitudes [mean ± standard error (s.e.); $n = 44$ to 134]. Each *marker* indicates different average loads ($F_0 = -1$ to $-5$ pN). The horizontal axis is normalized to the standard deviation (s.d.) of the noise, $\langle F_n^2 \rangle^{1/2}$. (d) Relative velocities of the probe using the same conditions as (c). The relative velocities were calculated as the velocity divided by the velocity without noise at the same load. (e) Numerical simulations for the velocity of the probe under semi-truncated Lévy noise of various magnitudes (*marker*, mean ± s.d.; $n = 10$). (f) Relative velocities of the data in (e). *Solid Lines* indicate theoretical predictions from Eqs. (S5) and (S8) in [27] (See also Appendix B).

The results so far show the effect of *white* noise applied to the probe. It is known, however, that actual intracellular fluctuations have a large frequency dependency [11,12]. To investigate the frequency dependency of the kinesin acceleration, we applied sinusoidal noise oscillating at different frequencies with a larger amplitude than the linear response range [21] and measured the average velocity. The acceleration showed a characteristic dependency on the frequency that peaked at ~200 Hz [Fig. 4(a)]. This tendency was also reproduced by numerical simulations [Fig. 4(b)]. These results indicate that the noise-induced acceleration strongly depends on its frequency characteristics.

Kinesin transports sub-μm sized vesicles or larger organelles within cells. In this study, instead of cargoes, force fluctuations were applied to the probe particle and indirectly transmitted to kinesin via the elastic linker. Thus, rapid fluctuations were attenuated due to the slow response of the probe. This attenuation can be explained by the Langevin model of the probe [Fig. 1(c) and Appendix B in [27]]. Simulations of the acceleration applying a simple Gaussian noise did not fit well with the theoretical prediction from the kinetic kinesin model but did agree with the prediction considering the response of the probe (Fig. S3 in [27]). Surprisingly, the theoretical prediction with simple Gaussian noise can quantitatively explain the simulations that applied semi-truncated Lévy noise (Fig. 2, *solid lines*, Eqs. (S5) and (S8) in [27]) without parameter modifications. In addition, we performed the velocity measurements by applying semi-truncated Gaussian noise, which was also quantitatively explained by the same prediction (Fig. S4 in [27]). These results suggest that the acceleration of kinesin is mainly determined by the magnitude (s.d.) of the applied noise, but it is insensitive to the shape of the noise distribution, at least under our experimental conditions.

The acceleration under sinusoidal noise tended to decrease even at lower frequencies [Fig. 4(b)], where the probe fluctuation is sufficiently transmitted to kinesin. The attenuation at low frequencies can be explained based on the kinetic model of kinesin [Fig. 1(b), Appendix C in [27]]. The perturbation expansion of the theory predicts that the acceleration has a similar characteristic frequency to the linear response as kinesin movements to the applied external force [21]. By estimating the high- and low-frequency limits of kinesin velocity, we quantitatively explained the frequency dependency of the kinetic kinesin model (Fig. S5 in [27]). Taken together, the ~200 Hz peak of the observed acceleration was quantitatively explained [Fig. 4(b), *lines*].

In our experiments, external forces were applied only to the direction of pulling backward; no force was applied in the forward direction due to the experimental limitations (Methods in [27]). The simulations applying noises without the limitations revealed that no remarkable acceleration was observed under the condition at which the average load is zero [Fig. 3(a) and Fig. S6 in [27], *open circles*]. In contrast, it was observed that another microtubule-based motor, dynein, moves faster when experiencing an external force shaking back and forth [18]. However, simulations with the same situation indicated that kinesin slows down in response to low-frequency rectangular pulse oscillations (Fig. S7 in [27]). The contrasting behaviors of the two motor proteins can be explained by the same mechanism, i.e. averaging the force-velocity relationship at forward and backward constant loads [18,27], indicating that the acceleration due to the slow back-and-forth oscillation is not a universal property but only occurs with a concave shape force-velocity relationship. On the other hand, in this letter, the acceleration of kinesin was observed under physiologically plausible fluctuations, where the total force was always applied backward and never directly forward. The different behaviors suggest that the noise-induced acceleration found in kinesin has a distinct mechanism from that previously found in dynein [18]. Because intracellular kinesins transporting cargoes are always enduring viscous drag, our application of noise in addition to the hindering load could better represent kinesins in the working environment.

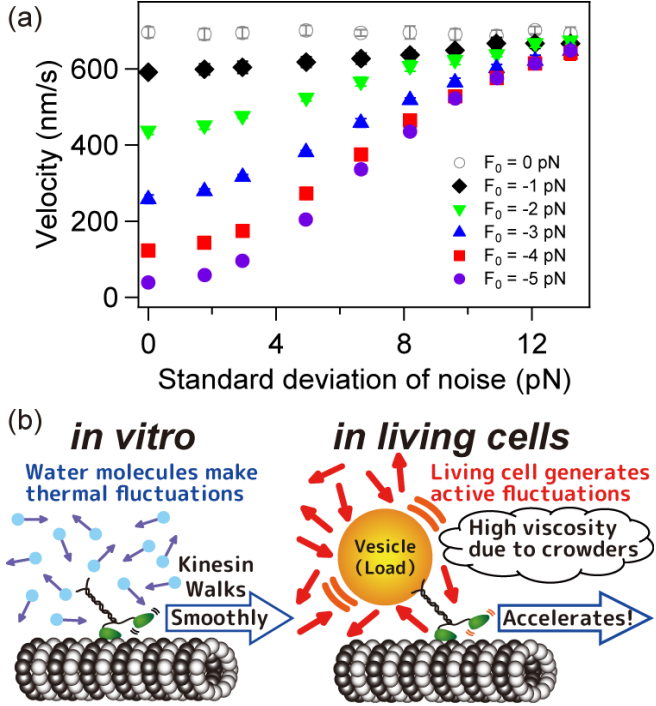

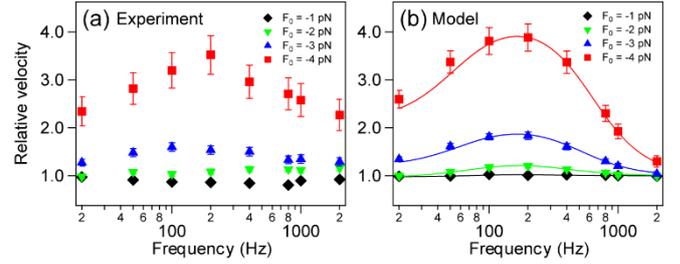

FIG. 3. (a) Numerical simulations for the velocities of the probe under semi-truncated Lévy noise limited to ±30 pN (i.e. physiological forces that exist in living cells [29]; mean ± s.d.; $n$ = 10). (b) Schematic drawing of the physiological implications of the noise-induced acceleration of kinesin. Kinesin utilizes thermal fluctuations to walk smoothly *in vitro* (*left*). When kinesin carries a vesicle within a living cell, it is expected that the kinesin utilizes the non-thermal fluctuations actively produced by the cell to achieve the same velocity as at no load even when intracellular crowders generate high viscosity (*right*).

FIG. 4. (a) The relative velocities of the probe particles under sinusoidal noise of various frequencies (mean ± s.e; $n$ = 39 to 173). An external force, $F_p = F_0 + A\sin(ft)$, was applied as a constant force (load), $F_0$, plus zero-mean sinusoidal forces with an amplitude of $A$, where the amplitude of the sinusoidal noise is the same as the load ($A = F_0$) at different frequencies, $f$. Different *markers* indicate different average loads ($F_0$ = 0 to -4 pN). (b) Numerical simulations for the relative velocities of the probe particles under sinusoidal noise of various frequencies (*markers*, mean ± s.d.; $n$ = 10). *Solid Lines* indicate the theoretical predictions from Eq. (S13) in Methods in [27] (See also Appendix C).

Although its acceleration response due to the low-frequency oscillation depends on the shape of the F-V relationship, dynein can also be expected to perform the same acceleration phenomenon observed in kinesin if active fluctuations are applied in the same relatively high-frequency range as used here and in intracellular conditions. Regardless of the details of the mathematical model, i.e. the shape of the graph combining each transition process, which depends on the kinetic scheme of the motor and generally differs between molecular motors, the external force dependence of each transition process is predicted to behave with common Arrhenius-type force dependency (Eq. 1 and as used in [18]). Therefore, the acceleration due to Jensen's inequality can always be predicted. Moreover, not only these molecular motors, but general enzymes also act with structural changes and obey the same Arrhenius equation. Thus, the noise-induced acceleration found in kinesin can be extrapolated to any molecular machine working in living cells.

So far, we have investigated only the steady-state properties, where kinesins are always moving along microtubules processively. Actual kinesin, however, stochastically dissociates from the microtubules after ~100 steps [31]. This dissociation becomes significantly faster when pulled forward than when pulled backward [32]. Thus, the addition of external forces in the forward direction (assisting force) makes the kinesin-tethered cargo dissociate rapidly from the microtubules. This property found in kinesin is completely different from full-length dynein dimers, which accelerate in response to the assisting force, but resembles the behavior of monomeric dynein [18]. Unlike Feynman's ratchet-type dynein, kinesin may save energy by rapidly dissociating from the microtubule in response to assisting forces, where the cargo is likely to be passively carried by the intracellular active fluctuations.

In an actual living cell, the high viscosity is realized by molecular crowding [33,34]. Thus, the result in Fig. 3a indicates that the unloaded velocity measured at low viscosity can be achieved by using active fluctuations even under crowding conditions [Fig. 3(b)]. However, intracellular crowders affect not only the movement of cargo through the viscosity but also the diffusion of kinesin's individual motor domains [35], meaning that the presence of crowders *per se* changes the motor's activity. Moreover, whereas the crowding state within cells is dense enough to freeze the dynamics due to a glass transition, a living cell's finite viscosity is achieved by fluidizing through active fluctuations [36]. Therefore, only considering the effect of high viscosity is insufficient to discuss motor activity within cells. Thus, the next challenge is to develop *in vitro* conditions that mimic the intracellular crowding environment with abundant crowders that have active fluctuations.

In summary, we show that an active fluctuating force accelerates the movement of single kinesin molecules, especially under high loads, suggesting that kinesin can move fast even under intracellular crowding conditions. Because the active fluctuations in cells are generated by the metabolic activity of a large number of motor molecules, the fluctuations *per se* consume large energies. Thus, the efficiency of each accelerated kinesin is apparently very low. Instead of the efficiency, as mentioned at the beginning of this letter, a different quantitative measure of the optimization, or fitness, for their working environment is desired. In addition, the acceleration is quantitatively explained by two universal theories, Jensen's inequality and the Arrhenius equation. Because general enzymes obey the same Arrhenius equation, intracellular active fluctuations should not be thought as just noise but as a sort of "vitality of life" that improves molecular activities in general.

One unresolved problem is that the velocities of vesicles transported by kinesin in cells are much faster than those observed *in vitro* [37]. Other factors have been implicated in vesicle transport within living cells, and some have been indicated to accelerate kinesin movement [38]. Our bottom-up approach, which approximates the *in vitro* assay to the intracellular active environment and gives an analysis that utilizes universal theories, sheds light on the physical/physiological principles underlying the use of active fluctuations by general biomolecules.

We thank Prof. Mutsunori Shirai and Prof. Makoto Furutani-Seiki for their kind permission for us to conduct our own research; Prof. Sin-ichi Sasa, Prof. Masahiro Ueda, and Prof. Shuji Ishihara for valuable discussions; and Dr. Peter Karagiannis for critically revising the manuscript. This work was supported by JSPS KAKENHI (JP15K05248, JP18K03564, JP19H05398, JP20H05535, and 21H00405 to TA and JP20H05536 to DM).

---

*ariga@yamaguchi-u.ac.jp

# SUPPLEMENTAL MATERIAL
# "Noise-induced acceleration of single molecule kinesin-1"

Takayuki Ariga, Keito Tateishi, Michio Tomishige, and Daisuke Mizuno

## I. METHODS

### A. Materials.

Tail-truncated human kinesin-1 cysteine-light mutant was used as "kinesin" in this study. Because the tail domain inhibits the activity by interacting with the catalytic heads [1], we genetically truncated the tail domain and instead added a six-histidine tag to the carboxy-terminal of the kinesin construct for attaching the probe beads and for the purifications. Microtubules were fluorescently labeled by ATTO 532 dye (Atto-tec, GmbH) for fluorescent imaging [2]. Protein expression and purification were carried out as previously described [3]. Probe beads (diameter, 520 ± 10 nm; Polysciences Inc.) were also fluorescently labeled and covalently conjugated with a 6xHistidine monoclonal antibody (9F2, Wako) [2]. The probe beads and kinesins were mixed on ice on the day before observation at a ratio of not more than one such that the probability of the movement was limited to less than 30% to ensure single-molecule conditions.

### B. Optical tweezers microscopy.

To apply actively fluctuating external forces to a single-molecule walking kinesin, a field-programmable gate array (FPGA)-based feedback-control optical tweezers microscope was utilized [2]. The probe bead attached to a single-molecule kinesin was trapped by a focused infrared laser (1064 nm, BL-106-TU-E, Spectra-Physics) and put on the microtubule rail by observing epi-fluorescent microscopy illuminated with a green laser (532 nm, DJ532-10, Thorlabs). The movement of the probe bead was acquired via quadrant photodiodes (QPD; S4349, Hamamatsu Photonics) by bright field imaging with an LED light (M660L3, Thorlabs) while applying desired external forces by controlling the trap distance via acousto-optic deflectors (AOD; DTD-274HD6M, IntraAction). The deflection angles of the AODs were controlled by two analog RF drivers (DE-272JM, IntraAction), to which ±10 V output signals from the FPGA board were input as ±1 V control signals after being multiplied by 1/10 with a self-made operational amplifier circuit. Because the electric circuits of the FPGA board can be changed just by programming, this device enables us to apply any type of external force to the probe particle following the kinesin walking. However, to achieve fast and precise control of the external force, the calibrated range of the QPD detection is limited to about ±210 nm, and the maximum output force (~10 pN) and the linear range (~200 nm) of the optical tweezers are also limited. These limitations provide constraints for our experiments.

The calibration for the displacement was performed with trapped beads by two-dimensional scanning around a 420 × 420 nm squared area with fifth-order polynomial fitting [4]. Previously, the linear response was measured by applying a sinusoidal perturbation force that was small enough not to affect the mean velocity of the kinesin movement [2]. In this study, we measured the change of the velocity by applying the perturbation forces largely exceeding the linear response range. Equipment manipulation and data acquisition were performed with custom-written LabVIEW programs (LabView 2018, National instruments). The probe and trap position data were recorded on a PC at a 20 kHz sampling rate.

### C. Generating active fluctuations.

The characteristic function of the symmetric (Lévy-type) stable distribution, $\varphi(z)$, can be written as

$$\varphi(z) = \exp\left(i\delta z - \gamma |z|^\alpha\right), \quad (S1)$$

where $\alpha$ is a stability parameter that determines the power law of the asymptotic tail as $\sim 1/|x|^{1+\alpha}$, $\delta$ is a location parameter, and $\gamma$ is a scale parameter. Here, the skewness parameter and the related term are omitted because of the symmetry. $\alpha$ is restricted to $0 < \alpha \leq 2$. When $\alpha$ is 2, the distribution shows a Gaussian distribution with a finite standard deviation; otherwise, the standard deviation diverges. In this study, we used $\delta = 0$ for zero mean and $\alpha = 1.5$ to mimic intracellular environments [5]. This type of stable distribution is also known as the Holtsmark distribution [6]. Using Eq. (S1), the distribution, $P(x)$, can be written as

$$P(x) = \frac{1}{2\pi} \int_{-\infty}^{\infty} \varphi(z) e^{-ixz} dz. \quad (S2)$$

Here, we numerically generated Lévy-type fluctuations according to an algorithm developed by Chambers et al. [7]. The fluctuating random variable with Lévy distributions, $S(\alpha, \gamma)$, was generated using two basic random variables as

$$S(\alpha, \gamma) = \gamma \times \frac{\sin(\alpha U)}{\cos(U)^{1/\alpha}} \left\{ \frac{\cos[U(1-\alpha)]}{E} \right\}^{\frac{1-\alpha}{\alpha}}, \quad (S3)$$

where $U$ is uniform noise on $(-\pi/2 \sim \pi/2)$, and $E$ is standard exponential noise. The fluctuations were applied as mean zero perturbation forces (noise) in addition to a constant external force (load) by changing the distance between the probe and trap position of the optical tweezers (trap distance). Due to the experimental limitation of the optical tweezers, a limited fluctuation of the trap distance was applied within the range of 100 ± 100 nm, where the trap stiffness is almost constant. Whereas the Lévy distribution in the mathematical form with Eqs. (S1) and (S2) contains infinitely large forces, real physical systems exhibit a finite range of active fluctuations [5,8-11] known as "Truncated-Lévy distributions". We truncated the Lévy distribution with the trap distance range by replacing the distance above the limit with an upper (or lower) limit. We call this modification a "semi-truncated Lévy distribution". The trap stiffness was adjusted using the equipartition law or Lorentzian fitting to the power spectrum density of the trace [2,12] such that the stiffness was set to the average load, $F_0$, divided by 100 nm. In this study, the average load was used in the range of $F_0 = -1$ pN to $-5$ pN, where the range of the noise is limited to ±1 pN to ±5 pN, respectively. Gaussian noise was also numerically generated and truncated within the experimental limitation range, called semi-truncated Gaussian noise. The semi-truncated Lévy noise generated by Eq. (S3) and the semi-truncated Gaussian noise are *white* noise, which does not have a frequency dependency. Thus, to investigate the frequency dependency, a sinusoidal fluctuating force, $F_n = A\sin(ft)$, where $A$ is the amplitude and $f$ is the frequency, was also applied as an external noise. All noises were applied at the same update rate as the sampling rate (20 kHz).

**D. Sample preparation.**

Detailed experimental procedures are described elsewhere [2]. Briefly, fluorescent microtubules were non-specifically attached onto a plasma-cleaned glass chamber. After removing the excess microtubules by infusing 1 mg/ml casein solution, the assay solution containing ~1 pM kinesin-coated beads was injected. All experiments were performed using an assay solution that contained 12 mM 1,4-piperazinediethanesulfonic acid (PIPES)-KOH (pH 6.8), 2 mM $MgCl_2$, 1 mM ethylene glycol tetraacetic acid (EGTA), 29 mM potassium acetate, 50 U/ml glucose oxidase, 50 U/ml catalase, 4.5 mg/ml glucose, 0.5% 2-mercaptoethanol, 0.4 mg/ml casein and 20 μM paclitaxel under nucleotide concentrations close to the physiological conditions inside cells (1 mM ATP, 0.1 mM ADP, and 1 mM potassium phosphoric acid; $\Delta\mu$ = 85 pN nm) at 25 ± 1 °C.

**E. Velocity analysis.**

The velocities of the probe movements were obtained from the slope of each trajectory of the time-series position data by fitting a linear function. When analyzing the velocity with sinusoidal noises, the fitting range was limited by adjusting the sine wave period such that the mean value of the sinusoidal noise was zero. Semi-truncated Lévy noise was applied by changing the trap distance with a scale parameter, $\gamma$. The magnitude of noise (s.d. of the applied forces) for each scale parameter was calculated by multiplying the trap stiffness and the s.d. of the trap distance. Whereas the s.d. for each trajectory was also distributed due to the limited number of noise data points, the s.d. for each scale parameter was calculated from a sufficiently large number of noise data points generated by the numerical simulations for normalization in Fig. 2 in the main text. Similarly, in the case of Gaussian noise, the s.d. for normalization in the figures were generated by the simulations but are not actually given. Trajectories containing long dwell times, which may be due to nonspecific adsorption, and instantaneous large position changes (much larger than 16 nm), which may be due to the detachment and re-attachment cycle, were omitted from the analysis. All data analysis was performed using Igor Pro 8.0 (Wavemetrics, Inc.).

**F. Numerical simulations.**

Numerical simulations with the mathematical kinesin model are based on our previous study [2]. Briefly, the kinesin motor is modeled as a jumping point with 8 nm steps that includes two internal kinetic states (Fig. 1b in the main text). The transition between the two states contains a force-independent rate, $k_c$, and two force-dependent transitions, $k_f$ and $k_b$, that obey Arrhenius-type force dependence:

$$k_{\{f,b\}}(F_m) = k_{\{f,b\}}^0 \exp\left(\frac{d_{\{f,b\}}F_m}{k_B T}\right), \tag{S4}$$

where $k_{\{f,b\}}^0$ is the rate constant at zero force, $d_{\{f,b\}}$ is the characteristic distance, $k_B$ is the Boltzmann constant, $T$ is the absolute temperature, and $F_m$ is the force applied to the motor. The subscripts $f$ and $b$ indicate forward and backward steps, respectively. From the model, the mean velocity, $v$, under a constant load is theoretically calculated as [2]

$$\langle v \rangle = d \times \frac{(k_f - k_b)k_c}{k_f + k_b + k_c}, \tag{S5}$$

where $\langle \ \rangle$ denotes the average and $d$ is the step size of kinesin ($d$ = 8 nm). The five parameters in the kinesin model are determined by fitting Eq. (S5) to the force-velocity relationship (Fig. S1a).

Note that the two-state model (Fig. 1b in the main text) is a simplified version of the current kinetic scheme of kinesin [13,14], which originally has many states. Since this simplification corresponds to the extraction of only the rate-limiting processes from many elementary processes in the kinetic scheme, increasing the number of states corresponds to the incorporation of non-rate-limiting processes into the model. In this case, even if the rates of these non-rate-limiting processes change due to the influence of external force fluctuations, the overall average velocity is not expected to be affected much, because the effect of the slowest rate-limiting process is dominant for the average velocity.

The probe's dynamics (Fig. 1c in the main text) is described by the Langevin equation as

$$\Gamma \frac{d}{dt}x_p = K(x_m - x_p) + F_0 + F_n + \xi, \tag{S6}$$

where $\Gamma$ is the viscous drag, $K$ is the spring constant of the stalk, and $x_p$ and $x_m$ are the position of the probe and motor,

respectively. The external force to the probe, $F_p = F_n + F_0$, contains a constant force (load), $F_0$, and a fluctuating force with a mean zero (noise), $F_n$. $\xi$ is white Gaussian thermal fluctuations that satisfy $\langle \xi \rangle = 0$ and $\langle \xi(t)\xi(t') \rangle = 2k_BT\Gamma\delta(t-t')$, where $\delta(t)$ is the delta function. The parameters for Langevin dynamics, $\Gamma$ and $K$, were determined experimentally from the power spectrum density of the displacement of the probe, which was stably bound to the microtubule via a single kinesin molecule bound to a non-hydrolyzable nucleotide, AMP-PNP [2,12] (Fig. S2). Here, we used $\Gamma = 3.25 \times 10^{-5}$ pN/nm s and $K = 3.99, 6.85, 8.91, 11.6$, and $15.9 \times 10^{-2}$ pN/nm for $F_0 = -1$ to $-5$ pN, respectively.

It should be noted that the external force, $F_m$, in Eq. (S4) is defined as a force directly applied to the kinesin molecule, but in our measurements and simulation system, the external force is applied to the probe and indirectly transmitted to the motor via the elastic spring. Therefore, the thermal fluctuation of the probe was added to the kinesin molecule as an additional external force, causing a small difference between the theoretical prediction and the simulation results (Fig. S1b). To overcome this issue, we introduced correction factors as written in Appendix A:

$$C_{\{f,b\}} = \exp\left(\frac{(K \cdot d_{\{f,b\}}/2 - \Gamma\langle v \rangle)d_{\{f,b\}}}{k_BT}\right). \quad (S7)$$

When the simulations were performed, the corrected parameters of $k_f^0/C_f$ and $k_b^0/C_b$ were used for the kinetic rate constants at no load instead of $k_f^0$ and $k_b^0$, respectively. The requirement for the correction factors indicates that the thermal fluctuation on the probe also accelerates the kinesin movement slightly.

All numerical simulations were performed for 10 trials and 30 sec trajectories with $5 \times 10^{-6}$ s time steps using Igor Pro 8.0. The sampling rate of the position data and the update rate of the applied external force were the same 20 kHz used in the experiments.

### G. Theoretical predictions for the noise-induced acceleration.

By considering Jensen's inequality, the acceleration of the force-dependent rate constants is qualitatively explained. However, both rate constants for forward and backward steps ($k_f$ and $k_b$) are accelerated by inducing noise. Therefore, the direction of the acceleration is not uniquely determined. However, the characteristic distance for the backward direction, $d_b$, is known to be very small [15-17], and the rate constant for backward steps increases less than the rate constant for forward steps. Thus, the mean velocity is basically increased by adding noise.

Quantitatively, the observed acceleration is explained by the theoretical predictions based on our mathematical kinesin model (Appendix B). The accelerated force-dependent kinetic rates of the probe movement when applying a simple Gaussian noise, $F_n$, at average load, $F_0$, are given as

$$\langle k_{\{f,b\}}(F_0 + F_n) \rangle = \exp\left[\left(\frac{d_{\{f,b\}}}{k_BT}\right)^2 \bar{G}\langle F_n^2 \rangle\right] k_{\{f,b\}}(F_0), \quad (S8)$$

where $\bar{G} \equiv \langle (F_m - F_0)^2 \rangle / \langle (F_p - F_0)^2 \rangle$ is a transmission parameter that indicates how much the variance of the fluctuating force input to the probe is attenuated when the force is transmitted to the kinesin molecule via the elastic linker. The parameter, $\bar{G}$, can be calculated by

$$\bar{G} = \frac{f_c}{f_n}\text{Arctan}\left(\frac{f_c}{f_n}\right), \quad (S9)$$

where $f_c \equiv K/2\pi\Gamma$ is the corner frequency, and $f_n$ is the Nyquist frequency (half of the sampling rate). Although these predictions are based on the case that a simple (untruncated) Gaussian noise is applied, the theoretical velocities obtained by substituting Eqs. (S8) and (S9) into Eq. (S5) agreed well with the velocities under semi-truncated Lévy noise and semi-truncated Gaussian noise in our experimental conditions (Fig. 2e,f in the main text and Fig. S4e,f).

The frequency dependency of the noise-induced acceleration is also explained quantitatively with an assumption (Appendix C). The accelerated force-dependent kinetic parameters of the probe movement when applying a sinusoidal noise, $F_n = A\sin(ft)$, at sufficiently high frequencies are given as

$$\langle k_{\{f,b\}}(F_0 + F_n) \rangle = I_0\left(\frac{d_{\{f,b\}}}{k_BT}G(f)A\right)k_{\{f,b\}}(F_0), \quad (S10)$$

where $I_0$ is the zeroth-order modified Bessel function of the first kind, and $G(f)$ is the Lorentzian-type transmission function via the probe and the linker and given as

$$G(f) = \frac{f_c^2}{f_c^2 + f^2}. \quad (S11)$$

By substituting Eqs. (S10) and (S11) into the theoretical mean velocity (S5), the accelerated velocity at the high frequency limit, $v_{high}$, can be obtained. The velocity is quantitatively explained for the simulated velocities of the kinesin molecule when applying 2 kHz sinusoidal noise of various amplitudes (Fig. S5a). On the other hand, when the frequency $f$ is extremely low (2.5 Hz), the velocity of the probe at low loads does not accelerate but slows (Fig. S5b). The velocity when applying sinusoidal noise at the low frequency limit was obtained simply by calculating the mean velocity from the force-velocity relationship (S5) as

$$v_{low} = \frac{1}{2\pi}\int_0^{2\pi} v(F_0 + A\sin\theta)d\theta. \quad (S12)$$

The theoretical prediction from Eq. (S12) agreed with the simulated velocities of the kinesin molecule when applying sinusoidal noise at 2.5 Hz (Fig. S5b l*ines*). The transition of the velocities between the high and low frequency limit can be described as

$$v = \frac{v_{low}(k_a/2\pi)^2 + v_{high}f^2}{(k_a/2\pi)^2 + f^2}. \quad (S13)$$

Here, we assumed that the characteristic frequency of the kinesin molecule response to the noise is $k_a/2\pi$, where $k_a \equiv k_f + k_b + k_c$. Although the assumption is based on the situation when a tiny perturbation force is applied, the theoretical prediction from Eq. (S13) agreed well with the frequency dependency of the noise-induced acceleration under our experimental conditions (Fig. 4c in the main text).

# II. SUPPLEMENTAL FIGURES

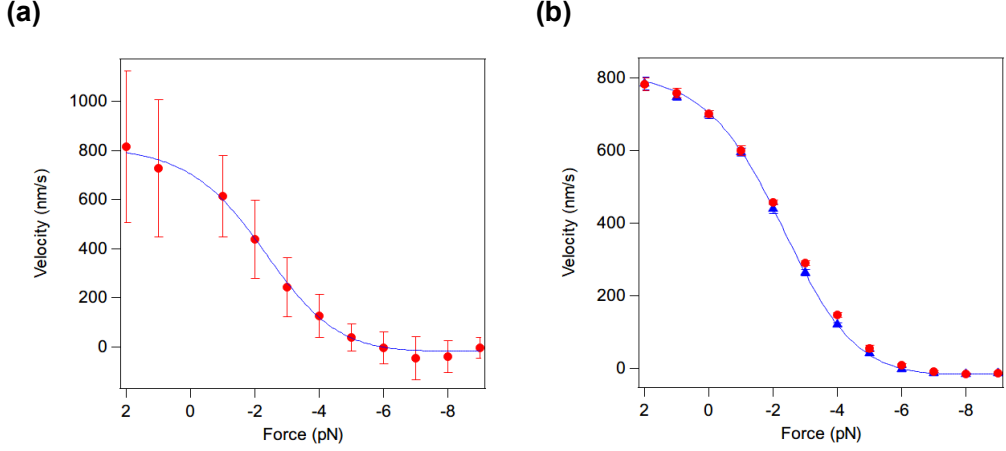

**FIG. S1. Force-velocity relationship of the kinesin movement.** (a) Force-velocity relationship of a probe pulled by a single kinesin molecule at 1 mM ATP, 0.1 mM ADP and 1 mM P$_i$ (mean ± s.d.; $n$ = 14 - 688). *Line* is fitted by the theoretical mean velocity with Eq. (A1) in Appendix A (Eq. (S5) in Methods). The fit parameters are $k_f^0$ = 1002 s$^{-1}$, $k_b^0$ = 27.9 s$^{-1}$, $k_c$ = 102 s$^{-1}$, $d_f$ = 3.61 nm, and $d_b$ = 1.14 nm. (b) Numerical simulations for the force-velocity relationship with and without correction factors (mean ± s.d., $n$ = 10). *Red circles* indicate the simulated velocity of the probe at different constant loads using the same kinetic parameters in (a) without corrections for thermal fluctuations. *Line* indicates the theoretical prediction using Eq. (A1), where the kinetic parameters are the same as in (a). Small deviations between the simulations (*circles*) and theoretical prediction (*line*) are observed. *Blue triangles* indicate the simulated velocities using parameter corrections for thermal fluctuations with Eqs. (A12) and (A13) in Appendix A (Eq. (S7) in Methods). The small deviations are thus eliminated.

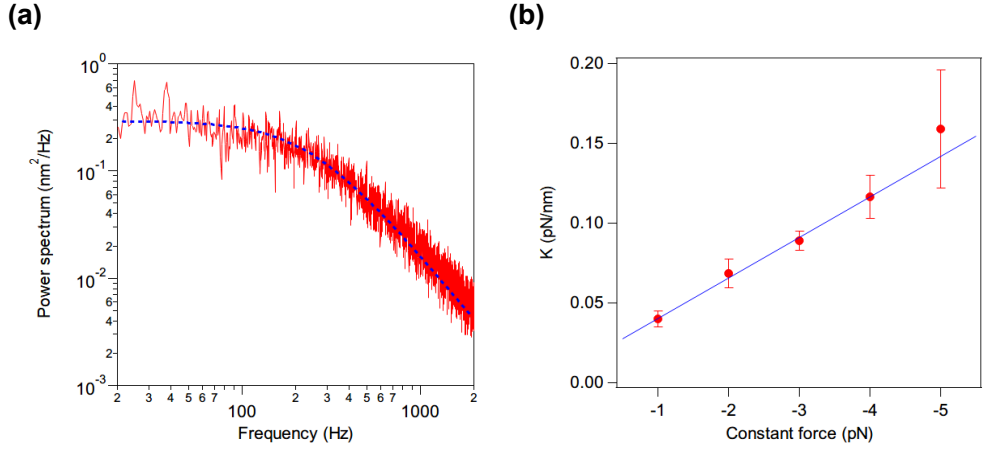

**FIG. S2. Calculating viscous drag Γ and spring constant *K*.** (a) Typical power spectrum density (PSD) of the probe position when the attached kinesin is stably bound to the microtubule at 1 mM AMP-PNP under -1 pN constant force. *Dashed line* indicates Lorentzian fitting with $PSD(f) = S_0/(1 + f^2/f_c^2)$, where $S_0$ is the horizontal line and $f_c$ is the corner frequency [12]. The spring constant was calculated as $K = 2k_B T/\pi S_0 f_c$, and the viscous drag was calculated as $\Gamma = k_B T/\pi^2 S_0 f_c^2$. (b) Relationship between the spring constant and external force (mean ± s.d.; *n* = 5). *Line* indicates linear fitting.

S6

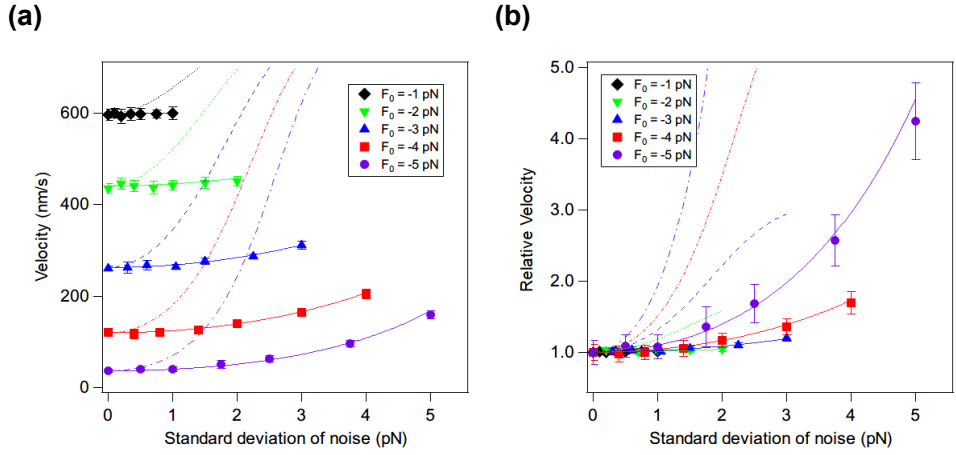

**FIG. S3. Numerical simulations and theoretical predictions for the probe under simple Gaussian noise.** (a) Velocities and (b) relative velocities of the probe under simple (untruncated) Gaussian noise. *Markers* indicate the simulation results at various average loads (mean ± s.d.; *n* = 10). *Dotted lines* indicate the theoretical predictions for the velocity of the kinesin molecule directly under Gaussian noise by using Eqs. (A1), (A15), and (A16) in Appendix B. *Solid lines* indicate theoretical predictions for the probe movement under the same noise by using Eqs. (A1) and (A24) while considering the transfer function derived from the Langevin dynamics of the probe and the linker. Although the *dotted lines* do not agree with the simulations, the *solid lines* fit well.

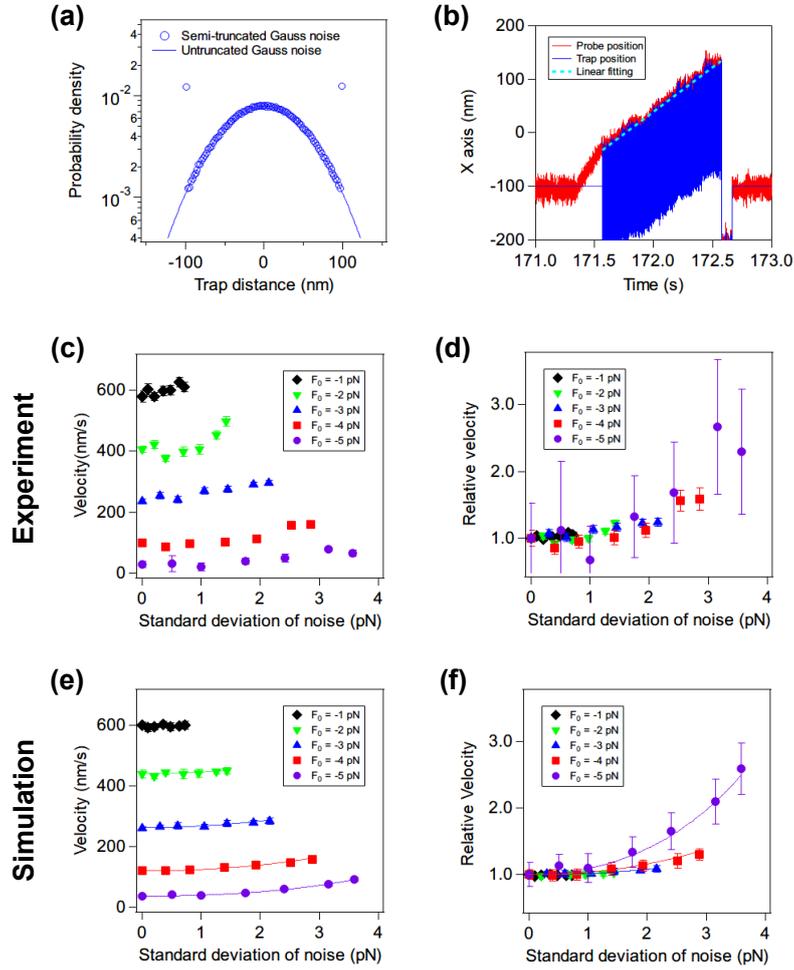

**FIG. S4. Experimental results and numerical simulations for kinesin movement under semi-truncated Gaussian noise.**
(a) The distribution of semi-truncated Gaussian noise (*marker*; s.d. = 50 nm). *Line* indicates a simple Gaussian distribution with the same s.d. (b) Typical trajectory of a probe particle pulled by a single kinesin molecule under semi-truncated Gaussian noise. The external force is applied as -4 pN average load and mean-zero Gaussian noise (s.d. = 50 nm), which truncated the trap distance to ±100 nm. (c) The velocity of kinesin beads under semi-truncated Gaussian noise of various s.d. (mean ± s.e.; $n$ = 45 to 159). Each *marker* indicates different average loads ($F_0$ = -1 to -5 pN). (d) The relative velocities of the data in (c). (e) Numerical simulation results for the velocities of the probe particles under semi-truncated Gaussian noise of various s.d. (*marker*; mean ± s.d.; $n$ = 10). (f) The relative velocities of the same data in (e). *Solid lines* indicate the theoretical prediction from Eqs. (A1) and (A24) in Appendix B [(S5) and (S8) in Methods].

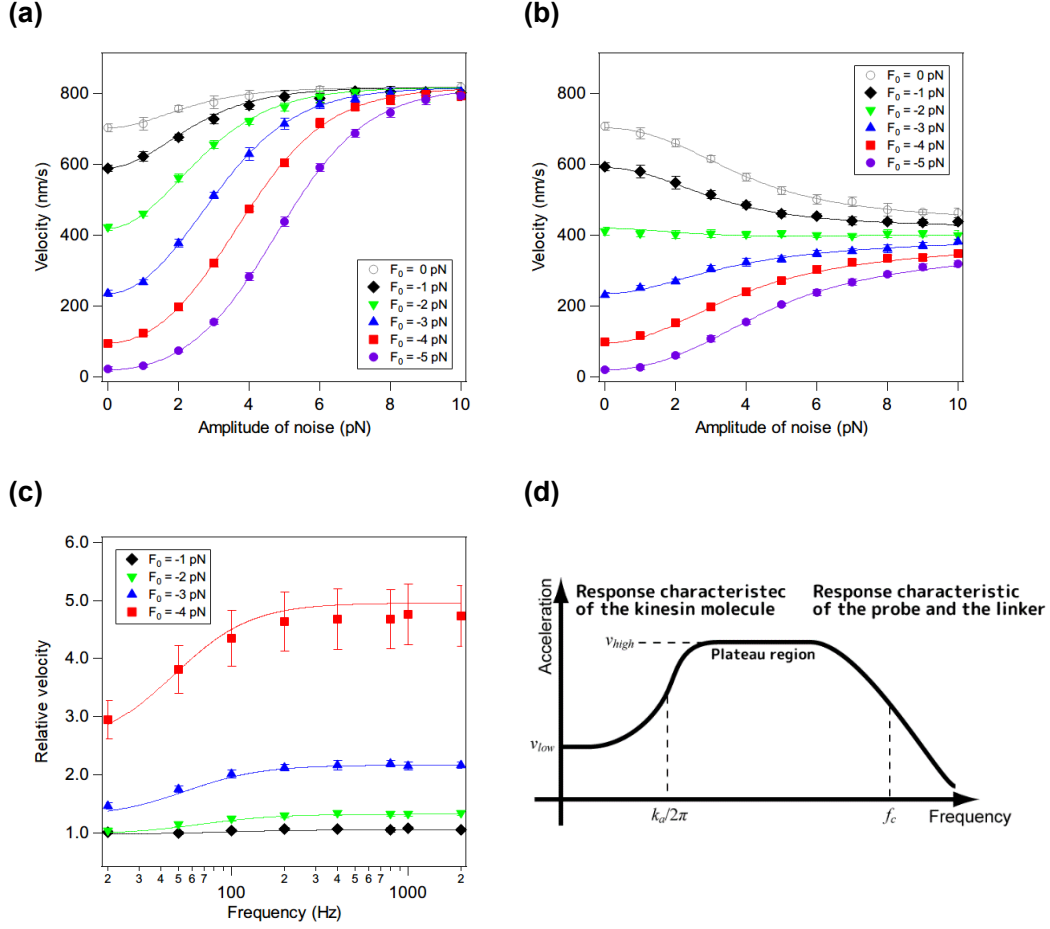

**FIG. S5. Theoretical predictions for the velocity of kinesin molecules directly under sinusoidal noise.** (a) The velocities of a kinesin molecule directly under untruncated sinusoidal noise, $F_n = A\sin(ft)$, where amplitude $A$ is variable and frequency $f$ = 2000 Hz. *Markers* indicate the numerical simulations for the velocities at various average loads ($F_0 = 0 \sim -5$ pN; mean ± s.d.; $n$ = 10). Langevin dynamics of the probe and linker are not used in the simulations. *Lines* indicate theoretical predictions for the kinesin velocity under sinusoidal noise at the high frequency limit, $v_{high}$, with Eqs. (A1) and (A28) in Appendix C. (b) The velocities of the kinesin molecule directly under untruncated sinusoidal noise at $f$ = 2.5 Hz. *Markers* indicate the numerical simulations for the velocities at various average loads ($F_0 = 0 \sim -5$ pN; mean ± s.d.; $n$ = 10). At low load conditions including no load (*open circle*: $F_0 = 0$ pN), the kinesin movement is not accelerated but slows down when low frequency sinusoidal noise is applied. *Lines* indicate the theoretical predictions for the kinesin velocity at the low frequency limit, $v_{low}$, with Eq. (A29) in Appendix C. Noting that in this situation 2.5 Hz noise is applied, almost no attenuation due to the probe's response occurs, and the simulation of the probe with the same noise application shows almost the same velocity profile (*data not shown*). (c) The relative velocities of the kinesin molecule directly under sinusoidal noise, where the amplitude is the same value as the average load ($A = F_0$). *Markers* indicate numerical simulations for the velocity at various average loads ($F_0 = -1 \sim -4$ pN; mean ± s.d.; $n$ = 10). *Lines* indicate theoretical predictions of the kinesin velocity with Eq. (A30). (d) Schematic drawing of the frequency characteristic for the noise-induced acceleration of kinesin. $k_a/2\pi$ is the characteristic frequency of the kinesin molecule, where $k_a \equiv k_f + k_b + k_c$. $f_c$ is the corner frequency ($f_c \equiv K/2\pi\Gamma$) of the probe. When these two characteristic frequencies are close, a single peak is observed.

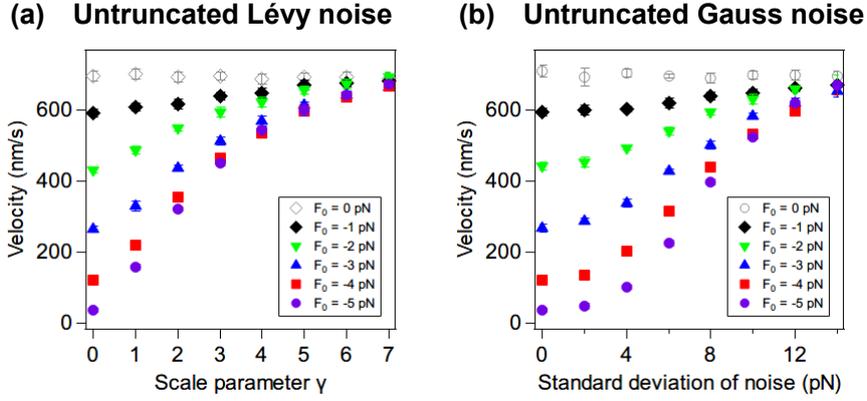

**FIG. S6. Numerical simulations for kinesin movement under active fluctuating forces without truncations.** (a) Numerical simulations of the velocities of the probe particles under untruncated Lévy noise of various scale parameters (mean ± s.d.; $n$ = 10). Note that the simulations with untruncated Lévy noise were not applied as the trap distance but directly applied as force using Eq. (S3) in Methods with scale parameter, $\gamma$, which has a pN scale. Because the s.d. of the Lévy noise intrinsically diverged, $\gamma$ is used for the horizontal axis. This result is very similar to that using a physiologically plausible constraint (truncated within ±30 pN; Fig. 3a in the main text), indicating that a remarkable acceleration response to extremely large and rare external forces in the untruncated noise (over ±30 pN) does not occur. However, such large and rare external forces significantly increase the average of the magnitude (s.d.) of the external force fluctuations, and the s.d. diverges at the mathematically ideal-type (untruncated) Lévy noise. Therefore, unlike in the experimental conditions, the acceleration under untruncated Lévy noise cannot be explained by the magnitude (s.d.) of the noise. (b) Simulations for the velocities of the probe under untruncated Gauss noise (mean ± s.d.; $n$ = 10).

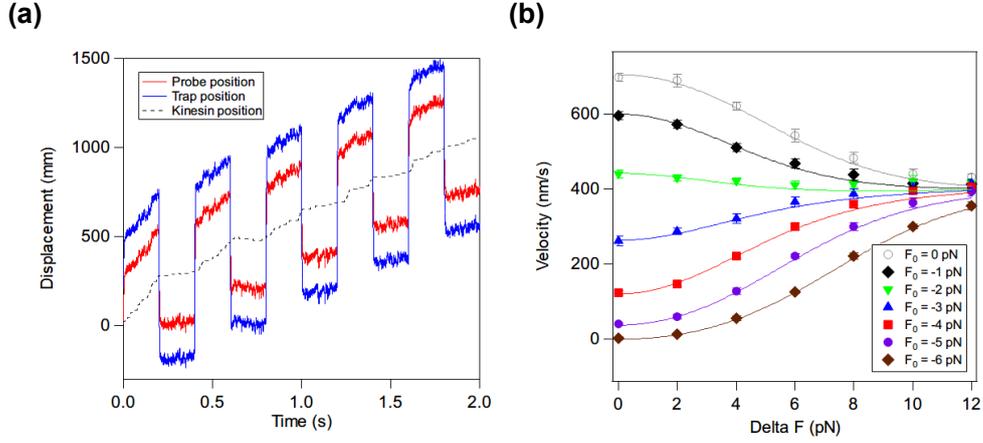

**FIG. S7. Numerical simulations for kinesin movement under rectangular pulse force oscillations.** (a) A typical simulated trajectory of a probe pulled by walking kinesin under a fluctuating force of rectangular pulse oscillations at conditions similar to a previous study on dynein [18]. The probe is driven with 2 pN assisting and -2 pN hindering forces at 2.5 Hz ($F_0$ = 0 pN and $\Delta F$ = 4 pN). (b) The velocity of the probe at various rectangular oscillating forces (mean ± s.d.; $n$ = 10). Each *marker* indicates the average loads ($F_0$ = 0 to -6 pN). *Lines* indicate the theoretical mean velocities, which were calculated by $\left[\langle v(F_0+\Delta F/2)\rangle+\langle v(F_0-\Delta F/2)\rangle\right]/2$ with Eq. (A1) in Appendix A (Eq. (S5) in Methods). The *lines* agree with the simulations. The velocity of the probe increases with oscillating forces at high loads, but decreases at low loads including no load (see also Fig. S5b). Noting that even under a load that kinesin moves backward ($F_0$ = -6 pN) without noise, it can be seen that kinesin moves forward when adding external force fluctuations.

# APPENDIX A: PARAMETER CORRECTIONS FOR THE ACCELERATING EFFECT FROM THERMAL FLUCTUATIONS

The theoretical formula for the mean velocity of the two-state kinesin model (Fig. 1b) was obtained in our previous study [2] as:

$$\langle v_m \rangle = d \times \frac{(k_f - k_b)k_c}{k_f + k_b + k_c}, \tag{A1}$$

where $\langle \ \rangle$ denotes the average, $d$ is the step size of kinesin, and $k_c$ is a force-independent kinetic parameter. Two kinetic parameters, $k_f$ and $k_b$, have an external force dependency as

$$k_{\{f,b\}}(F_m) = k^0_{\{f,b\}} \exp\left(\frac{F_m d_{\{f,b\}}}{k_B T}\right), \tag{A2}$$

where $F_m$ is an external force to the motor, $k_B$ is the Boltzmann constant, $T$ is the absolute temperature, $k_f^0$ and $k_b^0$ are the rate constants at zero load, and $d_f$ and $d_b$ are the characteristic distances for forward and backward steps, respectively. All parameters required for the numerical simulation can be obtained experimentally [2], of which the five parameters, $k_f^0$, $k_b^0$, $k_c$, $d_f$, and $d_b$, in the kinesin model were obtained by fitting the theoretical Eq. (A1) to the experimental results of the force-velocity (FV) relationship of the kinesin movement (Fig. S1a).

The theoretical Eqs. (A1) and (A2) (Eqs. (S4) and (S5) in Methods) are obtained as the average velocity of the kinesin movement when an external force, $F_m$, is directly applied to the kinesin molecule. However, in our measurement systems and numerical simulations, an external force, $F_p$, is applied to the probe particle and indirectly transmitted to the kinesin molecule through the elastic stalk region or linear spring (Fig. 1a and c in the main text). Therefore, when a simulation is performed using the parameters obtained just by fitting Eq. (A1) to the FV relationship of the motion of the probe particle, the simulated velocity of the probe, $\langle v_p \rangle$, deviates slightly from the theoretical prediction of the motor's mean velocity, $\langle v_m \rangle$ (Fig. S1b *circles* and *line*). This subtle difference is due to the difference between the force acting on the probe, $F_p$, and the force acting on the kinesin motor, $F_m$, where the thermal fluctuation of the probe is added to the kinesin molecule as an external force fluctuation, which is thought to cause a small acceleration of the kinesin movement.

To eliminate this discrepancy, correction factors were derived as follows.

Here, we consider a case in which active fluctuations (noise) are not added to external forces. The Langevin equation for the probe movement is given by:

$$\Gamma v_p = K(x_m - x_p) + \xi + F_p, \tag{A3}$$

where $x_m$ and $x_p$ are the positions of the kinesin and probe, respectively, $\Gamma$ is the viscous drag to the probe, $K$ is the spring constant of the elastic linker, and $\xi$ is the thermal fluctuation force that satisfies $\langle \xi \rangle = 0$ and $\langle \xi(t)\xi(t') \rangle = 2k_B T \Gamma \delta(t-t')$. The force on the kinesin molecule, $F_m$, is given by

$$F_m = -K(x_m - x_p). \tag{A4}$$

Using Eqs. (A3) and (A4), the difference between the external force applied to the probe and to the kinesin molecule, defined as $\delta F$, is expressed as

$$\delta F \equiv F_m - F_p = \xi - \Gamma v_p. \tag{A5}$$

The average of the rate constant, $k_f$, given the external force, $F_p$, is

$$\langle k_f \rangle = \left\langle k_f^0 \exp\left(\frac{F_m d_f}{k_B T}\right) \right\rangle$$
$$= \left\langle k_f^0 \exp\left(\frac{(F_p + \delta F)d_f}{k_B T}\right) \right\rangle \tag{A6}$$
$$= \left\langle \exp\left(\frac{\delta F d_f}{k_B T}\right) \right\rangle k_f^0 \exp\left(\frac{F_p d_f}{k_B T}\right).$$

The average rate can be written as the product of the correction factor,

$$C_f \equiv \left\langle \exp\left(\frac{\delta F d_f}{k_B T}\right) \right\rangle, \tag{A7}$$

and the term where $F_p$ is substituted as the external force in original Eq. (A2).

S12

From Eq. (A5), the mean and variance of $\delta F$ are given as

$$\langle \delta F \rangle = -\Gamma \langle v_p \rangle \tag{A8}$$

and

$$\langle (\delta F - \langle \delta F \rangle)^2 \rangle = \langle |\xi - \Gamma \delta v_p|^2 \rangle, \tag{A9}$$

respectively. Here, $\delta v_p$ is defined as a deviation from the average velocity of the probe.

Assuming that $\langle |\xi|^2 \rangle \gg \langle |\Gamma \delta v_p|^2 \rangle$ (this assumption is reasonable when the velocity fluctuation is observed [2]), the energy derived from the thermal fluctuation can be assumed to be equally distributed to the linear spring connecting kinesin and the probe. Thus, the equipartition law of the spring can be written as

$$\frac{1}{2} K \langle \delta x^2 \rangle = \frac{1}{2} k_B T, \tag{A10}$$

where $\langle \delta x^2 \rangle$ is defined as the variance of the extension of the spring. From Eqs. (A9) and (A10), we have

$$\langle (\delta F - \langle \delta F \rangle)^2 \rangle = K^2 \langle \delta x^2 \rangle \approx K \cdot k_B T. \tag{A11}$$

Since the thermal fluctuation is Gaussian, by using the average of (A8) and variance of (A9), the statistical average of the correction factor obtained by Eq. (A7) is expanded to the second order as

$$C_f \approx \exp\left( \frac{\langle \delta F \rangle d_f}{k_B T} + \frac{d_f^2 \langle (\delta F - \langle \delta F \rangle)^2 \rangle}{2(k_B T)^2} \right)$$

$$= \exp\left( \frac{(K \cdot d_f / 2 - \Gamma \langle v_p \rangle) d_f}{k_B T} \right). \tag{A12}$$

Similarly, the correction factor for $k_b$ is obtained as

$$C_b = \exp\left( \frac{(K \cdot d_b / 2 - \Gamma \langle v_p \rangle) d_b}{k_B T} \right). \tag{A13}$$

These correction factors can be obtained from experimentally accessible parameters.

When the simulation is performed, the corrected parameters obtained by dividing $k_f^0$ and $k_b^0$, which are obtained by fitting the theoretical equation to the probe's velocity, by the respective correction factors ($k_f^0/C_f$ and $k_b^0/C_b$) are used as the rate constants at no load (Fig. S1b *triangles*). It should be noted that the correction factors, $C_f$ and $C_b$, depend on the spring constant, $K$, and the mean velocity, $\langle v_p \rangle$, both of which depend on the external force.

# APPENDIX B: DERIVATION OF THE THEORETICAL FORMULA FOR WHITE NOISE APPLICATION

## 1. Fluctuating external force dependency of kinesin molecules

To quantitatively explain the noise-induced acceleration of kinesin experimentally observed in this study, here we will derive a theoretical formula to describe the noise-induced acceleration based on the mathematical kinesin model (Fig. 1b and c in the main text). Similar to the corrections for the thermal fluctuation (Appendix A), when zero-mean external force fluctuation (noise), $F_n$, is applied to the kinesin molecule, the average value of the rate constants for forward steps can be written as

$$\begin{aligned}\langle k_f(F_p)\rangle &= \left\langle k_f^0 \exp\left(\frac{F_p d_f}{k_B T}\right)\right\rangle \\ &= \left\langle k_f^0 \exp\left(\frac{(F_0 + F_n)d_f}{k_B T}\right)\right\rangle \\ &= \left\langle k_f^0 \exp\left(\frac{F_0 d_f}{k_B T}\right)\right\rangle\left\langle \exp\left(\frac{F_n d_f}{k_B T}\right)\right\rangle \\ &= \left\langle \exp\left(\frac{F_n d_f}{k_B T}\right)\right\rangle k_f(F_0). \end{aligned} \quad (A14)$$

Considering the convexity of the exponential function, the acceleration coefficient can be expected to be greater than 1 for any noise force, $F_n$, due to Jensen's inequality.

It should be noted that the rate constant for backward steps, $k_b$, is also accelerated by inducing noise. Therefore, the direction of acceleration is not uniquely determined by substituting the coefficients to Eq. (A1). However, because the characteristic distance for the backward direction, $d_b$, is known to be very small [15-17], the rate constant for backward steps increases at a smaller rate than it does for forward steps. Thus, it is qualitatively understood that the mean velocity is always increased by adding noise. Although it is difficult to determine the absolute value of the acceleration coefficient for general noise, this can be quantitatively determined using simple examples as follows.

## 2. Applying simple Gaussian noise to a kinesin molecule

First, we consider a system in which an external force, $F_m$, is given to a kinesin molecule by adding a constant external force (load), $F_0$, and white Gaussian noise, $F_n$, having an average of zero and variance of $\sigma^2$. By using the moment generating function of the probability distribution satisfying the Gaussian distribution, the acceleration coefficient in Eq. (A14) can be expanded as

$$\begin{aligned}\left\langle \exp\left(\frac{F_n d_f}{k_B T}\right)\right\rangle &= \exp\left[\frac{1}{2}\left(\frac{d_f}{k_B T}\right)^2 \langle F_n^2\rangle\right] \\ &= \exp\left[\frac{1}{2}\left(\frac{d_f}{k_B T}\right)^2 \sigma^2\right]. \end{aligned} \quad (A15)$$

Similarly, the coefficient for the back step is obtained as

$$\left\langle \exp\left(\frac{F_n d_b}{k_B T}\right)\right\rangle = \exp\left[\frac{1}{2}\left(\frac{d_b}{k_B T}\right)^2 \sigma^2\right]. \quad (A16)$$

Inserting Eqs. (A15) and (A16) into Eq. (A1) is expected to give the accelerated velocity when Gaussian noise is applied to the kinesin molecule. However, when the theoretical lines were plotted on the simulation result for the probe movement by applying simple Gaussian noise, they did not fit at all (Fig. S3 *dotted lines*).

## 3. Transfer function of the probe

The above results indicate the acceleration effect when the Gaussian fluctuating force is directly applied to the kinesin molecule. However, the actual external force is applied to the probe, not to the kinesin molecule, and the force is transferred to the kinesin molecule through an elastic linker modeled by a linear spring (Fig. 1c in the main text). Since the probe is embedded in the assay solution, the force applied to the probe is considered to attenuate due to the slow response of the probe. Thus, the transfer function of the force passing the probe and the linker is derived as follows.

The Langevin equation for the probe is given as

$$\Gamma \frac{d}{dt} x_p + K x_p = F_p. \quad (A17)$$

Here, to consider only the transfer function in a linear system consisting of a linear spring and a probe, the thermal fluctuations and the kinesin motion are neglected ($\xi = 0$ and $x_m = 0$). Using the Fourier transform, Eq. (A17) is written as

$$2\pi \Gamma (f_c - if) \tilde{x}_p = \tilde{F}_p, \quad (A18)$$

where ~ represents the Fourier transformed variables and $f_c \equiv K/2\pi\Gamma$ is the corner frequency of the system. By taking the square of both sides of Eq. (A18), the relation between the fluctuation of the probe and the external force is given as

$$K^2 |\tilde{x}_p|^2 = \frac{f_c^2}{(f_c^2 + f^2)} |\tilde{F}_p|^2. \quad (A19)$$

Because the force applied to the kinesin molecule, $F_m$, is given by Eq. (A4) and here the kinesin movement is neglected, by substituting the Fourier transformed relationship, $\tilde{F}_m = K\tilde{x}_p$, into Eq. (A19), we have

$$|\tilde{F}_m|^2 = \frac{f_c^2}{(f_c^2 + f^2)} |\tilde{F}_p|^2. \quad (A20)$$

Thus, we obtained the Lorentzian-type transfer function of the system consisting of the spring constant, $K$, and viscous resistance, $\Gamma$, as

$$G(f) = \frac{f_c^2}{(f_c^2 + f^2)}. \quad (A21)$$

Here, the transmission parameter, $\bar{G}$, is defined as

$$\bar{G} \equiv \frac{\langle (F_m - F_0)^2 \rangle}{\langle (F_p - F_0)^2 \rangle}, \quad (A22)$$

which describes how much the variance of the fluctuating force input to the probe is attenuated when the force is transmitted to the kinesin molecule. To obtain $\bar{G}$ with a Gaussian fluctuation of mean zero as an input, it is sufficient to obtain the variance in the time domain when a constant noise $|\tilde{F}_p|^2 = 1$ in the frequency domain is input. Because the noise is given as discrete time series data with a sampling rate of 20 kHz in both our experiments and simulations, by using Parseval's theorem, we have

$$\bar{G} = \int_0^{f_n} G(f) df = \int_0^{f_n} \frac{f_c^2}{f_c^2 + f} df = \frac{f_c}{f_n} \text{Arctan}\left(\frac{f_c}{f_n}\right), \quad (A23)$$

where $f_n$ is the Nyquist frequency (half of the sampling rate).

*4. Applying a simple Gaussian noise to the probe movement*

When a Gaussian fluctuation with a variance of $\langle (F_p - F_0)^2 \rangle = \langle F_n^2 \rangle = \sigma^2$ is applied to the probe, the variance of the external force on the kinesin molecule becomes $\langle (F_m - F_0)^2 \rangle = \bar{G}\sigma^2$. In this case, the average force-dependent rate constants, $k_f$ and $k_b$, are given as

$$\langle k_{\{f,b\}}(F_0 + F_n) \rangle = \exp\left[\frac{1}{2}\left(\frac{d_{\{f,b\}}}{k_B T}\right)^2 \bar{G}\sigma^2\right] k_{\{f,b\}}(F_0). \quad (A24)$$

By substituting Eq. (A24) into the theoretical mean velocity (A1), we obtain the accelerated mean velocity $\langle v_p(F_0 + F_n) \rangle$. This theoretically calculated mean velocity agreed well with the simulation results of the probe movement under a simple Gaussian fluctuating force (Fig. S3). The value of the transmission parameter, $\bar{G}$, varied from about 0.01 to 0.12 according to the change of the spring constant, $K$, which depends on the average external force. For example, when -5 pN average force is applied, the spring constant is $K = 0.159$ pN/nm, viscous drag is $\Gamma = 3.25 \times 10^{-5}$ pN/nm s, corner frequency is $f_c = 779$ Hz, and $\bar{G}$ is ~0.116.

Surprisingly, although the theoretical lines of the mean velocity according to Eq. (A24) were obtained by assuming simple Gaussian noise as the fluctuation of the external force, the same theoretical lines agreed well with the simulation results of the probe movement when applying non-Gaussian-shaped fluctuation forces, i.e. semi-truncated Lévy noise and semi-truncated Gaussian noise, without modifying the parameters (Fig. 2 in the main text and Fig. S4). This means that the acceleration phenomenon of kinesin can be quantitatively explained only by the variance (second moment) of the fluctuations regardless of the shape of the distribution, at least within the range of our experimentally constrained external force fluctuations.

# APPENDIX C: DERIVATION OF THE THEORETICAL FORMULA FOR SINUSOIDAL NOISE APPLICATIONS

### 1. Acceleration response of kinesin molecules by sinusoidal fluctuating force

Here we consider a sine wave fluctuation with frequency, $f$, and amplitude, $A$, as an external force,

$$F_n = A \sin(ft), \tag{A25}$$

applied to the two-state kinesin model (Fig. 1b in the main text). The acceleration coefficients for each force-dependent rate constant obtained by Eq. (A14) are given as

$$\left\langle \exp\left(\frac{d_{\{f,b\}} F_n}{k_B T}\right) \right\rangle = \left\langle \exp\left(\frac{d_{\{f,b\}} A}{k_B T} \sin ft\right) \right\rangle. \tag{A26}$$

In our experiments and simulations, the velocity was analyzed from the time series data by adjusting the sine wave period such that the average of the sinusoidal noise force is zero. Thus, the average of the coefficients can be obtained by dividing the integral of one period by the length of the period as follows:

$$\left\langle \exp\left[\frac{d_{\{f,b\}} A}{k_B T} \sin ft\right] \right\rangle = \frac{1}{2\pi} \int_0^{2\pi} \exp\left[\frac{d_{\{f,b\}} A}{k_B T} \sin \theta\right] d\theta$$
$$= I_0\left(\frac{d_{\{f,b\}} A}{k_B T}\right). \tag{A27}$$

Here, $I_0$ is the zeroth-order modified Bessel function of the first kind. The average force-dependent rates constants when applying sinusoidal noise with a sufficiently fast frequency (the meaning of this will be discussed later) are given as

$$\left\langle k_{\{f,b\}}(F_0 + F_n) \right\rangle = I_0\left(\frac{d_{\{f,b\}} A}{k_B T}\right) k_{\{f,b\}}(F_0). \tag{A28}$$

Substituting these rate constants into the theoretical equation for the mean velocity Eq. (A1) gives an accelerated velocity, $v_{high}$, when a sinusoidal noise is applied directly to the kinesin molecule. The theoretical lines from Eqs. (A1) and (A28) agree well with the simulations of the velocity of the kinesin molecule directly under the sinusoidal noise of sufficiently high frequency (2000 Hz) (Fig. S5a).

### 2. Velocity of kinesin under sinusoidal noise with a slow frequency limit

On the other hand, the simulation results obtained when the probe was shaken with rectangular wave noise of a slow alternation rate of 2.5 Hz were less at lower average loads (Fig. S7 *markers*). This is trivial, because it can be simply explained by the average of the forward and backward pulling velocities predicted from the FV relationship (Fig. S1 *lines*). Thus, it is expected that the velocity under sinusoidal noise at a low frequency limit can also be described by the average value of the velocity predicted from the FV relationship. Since the FV relationship is expressed by Eqs. (A1) and (A2), the velocity at the low frequency limit, $v_{low}$, can be obtained as the expected value obtained by simply substituting Eq. (A25) into these equations and averaging it along the range of the external force as

$$v_{low} \equiv \langle v(F_0 + A\sin ft)\rangle$$
$$= \frac{1}{2\pi}\int_0^{2\pi} v(F_0 + A\sin\theta)d\theta. \quad (A29)$$

The integral can be numerically calculated, and the theoretical lines from Eq. (A29) agree well with the magnitude dependency of the probe's velocity under sinusoidal noise of the sufficiently low frequency 2.5 Hz (Fig. S5b).

### 3. Kinesin's frequency dependency on sinusoidal noise

Eqs. (A28) and (A29) give the velocities of the high-frequency limit, $v_{high}$, and low-frequency limit, $v_{low}$, when the sinusoidal noise is applied to kinesin molecules. Neither limiting velocity depends on the frequency, but they do depend on the amplitude, $A$. The actual velocity of the kinesin molecule should transition between these two limiting values with some frequency dependency.

On the other hand, the frequency dependency of the velocity fluctuations (velocity correlation) of the kinesin molecule and the linear response function to tiny external forces have already been obtained in our previous study [2]. In both cases, the characteristic velocity, which is defined as the sum of the three rate constants, $k_a \equiv k_f + k_b + k_c$, exists, and the values for the response and correlation are proportional to $(\omega^2 + k_a^2)^{-1}$. (Note that similar to the derivation of the linear response function [2], we attempted a second order perturbation expansion and obtained a similar frequency dependency. However, this perturbation expansion is only valid in the range in which the external force fluctuation is sufficiently small. Thus, it cannot be used as a theoretical formula for the large external force fluctuations applied here.)

Therefore, it is reasonable to expect that the acceleration response to the external force fluctuation has a similar frequency dependency. Using this assumption, the velocity of the transition between $v_{high}$ and $v_{low}$ at the characteristic velocity $k_a$ is expressed as

$$v_m = \frac{v_{low}k_a^2 + v_{high}\omega^2}{k_a^2 + \omega^2}$$
$$= \frac{v_{low}(k_a/2\pi)^2 + v_{high}f^2}{(k_a/2\pi)^2 + f^2}. \quad (A30)$$

Despite the assumption being based on a tiny perturbation force, the theoretical lines from Eq. (A30) almost agree with the simulated velocities of a kinesin molecule to which a sinusoidal noise is directly applied (Fig. S5c).

### 4. Frequency dependency of the probe's velocity

As discussed in Appendix B, the fluctuating force applied to the kinesin molecule attenuates relative to the force applied to the probe through a Lorentzian-type transfer function (A21) having a corner frequency of $f_c = K/2\pi\Gamma$. Thus, by applying the transfer function (A21) to the sinusoidal external force in the average velocity of the kinesin molecule (A30), the theoretical lines quantitatively agree with the simulated velocity of the probe (Fig. 4b in the main text).

When the corner frequency $f_c$ is about 800 Hz, the probe's velocity at high frequencies begins to attenuate at around 200 Hz, explaining the frequency dependency on the high frequency side in the probe's velocity. On the other hand, the behavior on the low frequency side, where the force fluctuation is transmitted without being attenuated by the probe or the linker, can be approximated by introducing the characteristic frequency, $k_a/2\pi$, as expressed in Eq. (A30). Although the attenuation of the acceleration at the low frequency side and the high frequency side were derived from different mechanisms, a single peak at around 200 Hz was coincidentally observed (Fig. 4 in the main text).

In general, the frequency dependency of the noise-induced acceleration has a trapezoidal shape sandwiched between $k_a/2\pi$ and $f_c$ (Fig. S5d). Since $k_a/2\pi$ depends on $k_f$ and $k_b$, and $f_c$ depends on the spring constant $K$, both characteristic frequencies depend on the applied load. In addition, the characteristic frequency of the kinesin molecule, $k_a/2\pi$, changes according to the degree of acceleration; that is, it depends on the frequency and amplitude of the external force fluctuations. It should be noted that, when the external force fluctuation is too large, the theoretical prediction at intermediate frequencies fails. This is probably because the fluctuation response characteristic of the kinesin molecule expressed by Eq. (A30) is derived from the assumption obtained with sufficiently small perturbation forces (linear response range). Further nonlinear effects will need to be incorporated to explain quantitatively the acceleration at larger fluctuating forces than those used in our experimental conditions.

# SUPPLEMENTAL REFERENCES